\documentstyle[11pt,amssymb,newlfont]{amsart}
\newcommand{\R}{\Bbb R}
\newcommand{\N}{\Bbb N}
\newcommand{\diver}{\operatorname{div}}

\newcommand{\supp}{\operatorname{supp}}
\newcommand{\be}[1]{\begin{equation}\label{#1}}
\renewcommand{\phi}{\varphi}
\newcommand{\eps}{\varepsilon}

\newcommand{\bml}[1]{\begin{multline}\label{#1}}
\newcommand{\bes}{\begin{equation*}}
\newcommand{\bs}{\begin{split}}
\newcommand{\bdm}{\begin{displaymath}}
\newcommand{\edm}{\end{displaymath}}
\newcommand{\di}{\partial}
\renewcommand{\phi}{\varphi}
\newtheorem{th}{Theorem}[section]
\newtheorem{lem}{Lemma}[section]

\theoremstyle{definition} 
\theoremstyle{remark}

\begin{document}
\title[]{Flow of electrorheological fluids 
under the conditions of slip on the boundary.}

\author[]{R.H.W. Hoppe, W.G. Litvinov.}
\date{}
\subjclass{35Q35}
\address{\newline
Lehrstuhl f\"ur Angewandte Analysis mit Schwerpunkt Numerik
\hspace*{3ex}\newline
Universit\"at Augsburg
\newline Universit\"atsstrasse, 14
\newline 86159 Augsburg, Germany
\hspace*{3ex}\newline
E-mails:\newline
hoppe  math.uni-augsburg.de\newline
litvinov  math.uni-augsburg.de.} 

\begin{abstract}

We derive general conditions of slip of a fluid on the boundary. Under these
conditions the velocity of the fluid on
the immovable boundary is a function of the normal and tangential components of
the force acting on the surface of the fluid. A problem on stationary flow
 of an electrorheological fluid in which the terms of slip are specified
on one part of the boundary and surface forces are given on the other
is formulated and studied. Existence of a solution of this problem is proved
by using the methods of penalty functions, monotonicity and compactness.
It is shown that the method of penalty functions and the Galerkin 
approximations  can be used for the approximate
solution of the problem under consideration.
\end{abstract}
\maketitle

\makeatletter\@addtoreset{equation}{section}\makeatother
\def\theequation{\arabic{section}.\arabic{equation}}

\section{Introduction}
Electrorheological fluids are smart materials which
are concentrated suspensions of polarizable particles 
in a nonconducting dielectric liquid. In moderately large electric fields, the
particles form chains along the field lines, and these chains then aggregate to 
form columns \cite{9}. These chainlike and columnar 
structures cause dramatic changes in
the rheological properties of the suspensions. The fluids become anisotropic,
the apparent viscosity (the resistance to flow) in the direction orthogonal to
the direction of electric field abruptly increases, while the apparent viscosity
in the direction of the electric field changes not so drastically. 

The chainlike and columnar structures are destroyed under the action of large 
stresses, and then the apparent viscosity of the fluid decreases and the fluid
becomes less anisotropic.

On the basis of experimental results, the following constitutive equation
was developed in \cite{3}:
\begin{equation}\label{1.1}
\sigma_{ij}(p,u,E)=-p\delta_{ij}+2\phi(I(u),\vert E\vert,\mu(u,E))\eps_{ij}
(u), \quad i,j=1,\dots,n, \quad n=2 \mbox{ or } 3.
\end{equation}
Here, $\sigma_{ij}(p,u,E)$ are the components of the stress tensor which 
depend on the pressure $p$, the velocity vector $u=(u_1,\dots,u_n)$ and the 
electric field strength $E=(E_1,\dots,E_n)$, $\delta_{ij}$ are the components
of the unit tensor (the Kronecker delta), and 
$\eps_{ij}(u)$ are the components of the rate of strain tensor
\begin{equation}\label{1.2}
\eps_{ij}(u)=\frac12\big(\frac{\di u_i}{\di x_j}\,+\,\frac{\di u_j}{\di x_i}
\big).
\end{equation}
Moreover, $I(u)$ is the second invariant of the rate of strain tensor
\begin{equation}\label{1.3}
I(u)=\sum_{i,j=1}^n (\eps_{ij}(u))^2,
\end{equation}
and $\phi$ is the viscosity function depending on $I(u)$, $\vert E\vert$ and $\mu(
u,E)$, where
\begin{equation}\label{1.4}
(\mu(u,E))(x)=\Big(\frac{u(x)}{\vert u(x)\vert},\,\,\frac{E(x)}{\vert E(x)
\vert}\Big)_{\R^n}^2\,=\,\frac{(\sum_{i=1}^n u_i(x)E_i(x))^2}{(\sum_{i=1}^n
(u_i(x))^2)(\sum_{i=1}^n(E_i(x))^2)}.
\end{equation}
So $\mu(u,E)$ is the square of the scalar product of the unit vectors $\frac
{u}{\vert u\vert}$ and $\frac E{\vert E\vert}$. The function $\mu$ is defined
by \eqref{1.4} in the case of an immovable frame of reference. If the frame
of reference moves uniformly with a constant velocity $\check u=(\check u_1,
\dots,\check u_n)$, then we set:
\begin{equation}\label{1.5}
\mu(u,E)(x)=\Big(\frac{u(x)+\check u}{\vert u(x)+\check u\vert},\,\,\frac
{E(x)}{\vert E(x)\vert}\Big)_{\R^n}^2.
\end{equation}
As the scalar product of two vectors is independent of the frame of reference,
the constitutive equation \eqref{1.1} is invariant with respect to the group
of Galilei transformations of the frame of reference that are represented as
a product of time-independent translations, rotations and uniform motions.

The presence of the function $\mu$ in the constitutive equation \eqref{1.1}
is connected with the anisotropy of the electrorheological fluid under which
the viscosity of the fluid depends on the angle between the vectors of velocity and the vector of electric field. 

The function $\mu$ defined by \eqref{1.4}, \eqref{1.5} is not specified at
$E=0$ and at $u=0$, and there does not exist an extension by continuity 
to the values of $u=0$ and $E=0$. However, at $E=0$ there is no influence of 
the electric field, 
and the function $\mu(u,E)$ need not to be specified at $E=0$. Likewise, in case 
that the 
measure of the set of points $x$ at which $u(x)=0$ is zero, the function $\mu$ 
need not also be specified at $u=0$. But in the general the  function $\mu$
can be defined as follows:
\begin{equation}\label{1.7}
\mu(u,E)(x)=\Big(\frac{\alpha\tilde I+u(x)+\check u}{\alpha\sqrt n+\vert u(x)+
\check u\vert},\,\,\frac{E(x)}{\vert E(x)\vert}\Big)_{\R^n}^2,
\end{equation}
where $\tilde I$ denotes  a vector with components equal to one, and $\alpha$ 
is a small
positive constant. If $u(x)\ne 0$ almost everywhere in $\Omega$, one can choose
$\alpha=0$.

The viscosity function $\phi$ is identified by approximation of flow curves, 
see \cite{3}, and it was shown in \cite{3} that it can be represented as 
follows:

\begin{equation}\label{1.8}
\phi(I(u),\vert E\vert,\mu(u,E))=b(\vert E\vert,\mu(u,E))(\lambda+I(u))^
{-\frac12}+\psi(I(u),\vert E\vert,\mu(u,E)),
\end{equation}
where $\lambda$ is a small parameter, $\lambda\ge 0$.

The equations for the functions $E$ and $(p,v)$ are separated, (see \cite{3}).
Because of this, we assume here and thereafter that the function of electric
field $E$ is known. 

Various problems on stationary flow of electrorheological fluids under mixed 
boundary conditions such that velocities and surface forces are prescribed 
on different parts of the boundary are investigated in \cite{3}. This
formulation assumes that the fluid adheres to a hard boundary, that is the
velocity of the fluid on the hard boundary is equal to the velocity of the
boundary.

But at some conditions, wall effects appear, the velocity of a fluid on the 
hard boundary can be different from the velocity of the hard boundary. In
particular, hard particles of electrorheological suspensions may slip along
the hard boundary.

It was shown experimentally that magnetic suspensions, whose conduct is
similar to the conduct of electrorheological fluids, exhibit wall effect,
see \cite{5}. This effect depends both on the surface roughness of the wall
and on the force pressing the particles against the surface of the wall.

In Section 2, we derive the boundary conditions of slip. In Sections 3 and 4,
we formulate a boundary value problem on stationary flow of the 
electrorheological fluid under the condition of slip on the boundary and
present a theorem on the existence of a solution of this problem.  Sections
5 and 6 are devoted to the proof of the existence result and construction
of approximate solutions by using the method of penalty functions. In Section 7,
we show that Galerkin approximations can be used for approximate solution of our problem.

Since the constitutive equations of nonlinear viscous and viscous fluids
are partial cases of the equation \eqref{1.1}, the results presented in
this paper can be applied to nonlinear viscous and viscous fluids.

\section{Frictional force and the velocity of slip on a hard boundary.}

Let $\Omega\subset\R^n$ be a domain in which a fluid flows. Let $S$ be the
boundary of $\Omega$ and $S_1$ be a part of $S$ which corresponds to a hard 
immovable wall. We assume that the fluid slips on $S_1$. Let $F(s)=\sum
_{i=1}^n F_i(s)\zeta_i$ be an external surface force acting on the fluid.
Here $\zeta_i$ are unit vectors directed along the coordinate axes $x_i$,
$F_i$ scalar functions of points $s$ of $S_1$.

We represent the function $F$ in the form
\begin{equation}\label{2.1}
F(s)=F^\nu(s)+F^\tau(s), \qquad s\in S_1,
\end{equation}
where $F^\nu$ and $F^\tau$ are the normal and the tangential surface forces.
\begin{align}
&F^\nu(s)=F_\nu(s)\nu(s), \qquad  F_\nu(s)=\sum_{i=1}^n F_i(s)\nu_i(s),
                                                \label{2.2}\\
&F^\tau(s)=F(s)-F^\nu(s)=\sum_{i=1}^n F_{\tau i}(s)\zeta_i, \qquad
  F_{\tau i}(s)=F_i(s)-F_\nu(s)\nu_i(s),           \label{2.3}
\end{align}
where $\nu=(\nu_1,\dots,\nu_n)$ is the unit outward normal to $S_1$.

Analogously, the velocity vector $u$ on the boundary is represented in the
form
\begin{align}
&u(s)=\sum_{i=1}^n u_i(s)\zeta_i=u^\nu(s)+u^\tau(s),  \notag\\
&u^\nu(s)=u_\nu(s)\nu(s), \qquad u_\nu(s)=\sum_{i=1}^n u_i(s)\nu_i(s),
                                                      \notag\\
&u^\tau(s)=u(s)-u^\nu(s)=\sum_{i=1}^n u_{\tau i}(s)\zeta_i, \label{2.4}                                
\end{align}
where
\begin{equation}\label{2.5}
u_{\tau i}(s)=u_i(s)-u_\nu(s)\nu_i(s).
\end{equation}
We consider the following boundary conditions on $S_1$:
\begin{gather}
u_\nu(s)=0, \qquad s\in S_1,                           \label{2.6}\\
F^\tau(s)=-\chi(F_\nu(s),\vert u^\tau(s)\vert^2)u^\tau(s), \quad s\in S_1.
                                                       \label{2.7}
\end{gather}
Here $\chi$ is the function of slip that depends on the normal component
of the surface force $F_\nu$ and on the square of the module of the tangential
velocity $u^\tau$. 

Formula \eqref{2.7} is a generalization of Navier's condition of slip in which $\chi$ is a positive 
constant, the nonlinear modification of Navier's condition of slip  
in which $\chi$ is a function of $\vert u^\tau\vert$, and  Coulomb's law of friction in which $\chi=\infty$ at $\vert F^\tau \vert < c_1\vert F^\nu \vert$
and $\chi=c$ at $\vert F^\tau \vert=c_1\vert F^\nu \vert$, $c, c_1$ are positive constants.

We note that problems on flow of nonlinear viscous fluids in which $\chi$
is a function of $\vert u^\tau\vert$ were investigated  in \cite{8}. 

The function $\chi$ accepts positive values, $\chi$ does not depend 
of $F_\nu$
at $F_\nu>0$, and it rises as $F_\nu$ decreases. The sign minus in \eqref{2.7} designates that 
the velocity of slip of the fluid  is in opposition to
the tangential surface force, i.e. the frictional force is in
opposition to the direction of motion, and the module of the slip velocity
is equal to $\vert F^\tau(s)\vert(\chi(F_\nu(s),\vert u^\tau(s)\vert^2))
^{-1}$.

In the special case that $\chi(y_1,y_2)=\infty$ for an arbitrary $(y_1,
y_2)\in\R\times\R_+$, formulas \eqref{2.6}, \eqref{2.7} imply $u\Big\vert_
{S_1}=0$, i.e. the fluid adheres to the hard boundary, and in the case of
$\chi(y_1,y_2)=0$ for an arbitrary $(y_1,y_2)\in\R\times\R_+$,
 \eqref{2.7}
yields $F^\tau\Big\vert_{S_1}=0$, i.e. the frictional force is equal to zero.
The relation \eqref{2.6} designates that the fluid does not flow through
the hard wall $S_1$.

For the constitutive equation \eqref{1.1} the components $F_i$ of the
surface force $F=(F_1,\dots,F_n)$ are defined by
\begin{equation}\label{2.8}
F_i=[-p\delta_{ij}+2\phi(I(u),\vert E\vert,\mu(u,E))\eps_{ij}
(u)]\nu_j\Big\vert_{S_1}, \quad i=1,\dots,n,
\end{equation}
and the normal component $F_\nu$ of the surface force is determined by 
\eqref{2.2}. In \eqref{2.8}  and below the Einstein convention on summation 
over repeated index is applied.

Let $P$ be an operator of regularization given by
\begin{equation}\label{2.9}
Pv(x)=\int_{\R^n}\omega(\vert x-x'\vert)v(x')\,dx', \qquad x\in\overline\Omega,
\end{equation}
where
\begin{gather}
\omega\in C^\infty\,(\R_+), \quad \supp \omega \in[0,a], \quad \omega(z)\ge 0,
\quad z\in\R_+,                                \notag\\
\int_{\R^n}\omega(\vert x\vert)\,dx=1, \qquad  a \text{ is a small positive
constant}.                                     \label{2.10}
\end{gather}
Here, we assume that the function $v$ is extended to $\R^n$.

We denote by $F_{r\nu}(p,u)$ the normal component of the surface force
calculated by the regularized functions of pressure $p$ and velocity $u$.
According to \eqref{2.2} and \eqref{2.8}, the function $F_{r\nu}(p,u)$ is
defined as follows:
\begin{equation}\label{2.11}
F_{r\nu}(p,u)=[-Pp+2\phi(I(Pu),\vert E\vert,\mu(Pu,E))\eps_{ij}(Pu)\nu_i
\nu_j]\Big\vert_{S_1}.
\end{equation}
We change the function $F_\nu$ in \eqref{2.7} for the function $F_{r\nu}
(p,u)$. Then, we obtain the following boundary condition:
\begin{equation}\label{2.12}
F^\tau(s)=-\chi(F_{r\nu}(p,u)(s),\vert\, u^\tau(s)\vert^2)u^\tau(s),
\qquad s\in S_1.
\end{equation}  
From the physical point of view, \eqref{2.12} denotes that the model is not
local, the velocity of slip at a point $s\in S_1$ depends on the averaged 
normal surface force $F_{r\nu}(p,u)$ which in its turn is defined by the 
values of pressure and the derivatives of the velocity at points belonging
 to some small vicinity of the point $s$. This is natural from the physical
view-point.

Such nonlocal approach is also connected with the fact that the velocity
of slip depends on the surface roughness which is not a local characteristic.

Taking \eqref{2.2}, \eqref{2.3} and \eqref{2.4} into account, we represent
\eqref{2.12} in the following form:
\begin{equation}\label{2.13}
F_i-\Big(\sum_{k=1}^n F_k\nu_k\Big)\nu_i=-\chi\Big(F_{r\nu}(p,u),\,\sum_{k=1}
^n u_{\tau k}^2\Big)u_{\tau i} \quad \text{on }S_1,  \quad     
i=1,\dots,n.
\end{equation}
Finally, we obtain by \eqref{2.8} and \eqref{2.13} the following boundary
condition of slip:
\begin{gather}
2\phi(I(u),\vert E\vert,\mu(u,E))[\eps_{ij}(u)\nu_j-
\eps_{kj}(u)\nu_j\nu_k\nu_i]        
          =-\chi(F_{r\nu}(p,u),\,\sum_{k=1}^nu_{\tau k}^2)u_{\tau i}
            \text{ on } S_1,             \notag\\
\quad i=1,\dots,n,   \label{2.14}
\end{gather}
where $F_{r\nu}(p,u)$ is defined by \eqref{2.11}.

\section{Governing equations and assumptions.}
We consider stationary flow problem under the Stokes approximation, i.e. 
we ignore inertial forces which are assumed to be small as compared with
the internal forces caused by the viscous stresses. Then, the motion equations 
take the following form:
\begin{equation}\label{3.1}
\frac{\di p}{\di x_i}-2\,\frac{\di}{\di x_j}[\phi(I(u),\vert E\vert,\mu
    (u,E))\eps_{ij}(u)]=K_i \text{ in } \Omega, \quad i=1,\dots,n,
\end{equation}
where $K_i$ are the components of the volume force vector $K$.

The velocity function $u$ meets the incompressibility condition
\begin{equation}\label{3.2}
\diver u=\sum_{i=1}^n\frac{\di u_i}{\di x_i}=0 \quad \text{in } \Omega.
\end{equation}

We assume that $\Omega$ is a bounded domain in $\R^n$, $n=2$ or $3$
with a Lipschitz continuous boundary $S$. Suppose that $S_1$ and $S_2$
are open non-empty subsets of $S$ such that $S_1\cap S_2=\emptyset$, and
$\overline S_1\cup\overline S_2=S$. 

We consider mixed boundary conditions for the functions $u$, $p$. Wherein,
the terms of slip \eqref{2.6}, \eqref{2.14}  are specified on $S_1$ and
surface forces are given on $S_2$, i.e.
\begin{equation}\label{3.3}
[-p\delta_{ij}+2\phi(I(u),\vert E\vert,\mu(u,E))\eps_{ij}(u)]\nu_j\Big
\vert_{S_2}=F_i, \quad i=1,\dots,n.
\end{equation}
We assume that
\begin{description}
\item[(C1)]
               $\phi:(y_1,y_2,y_3)\to\phi(y_1,y_2,y_3)$ is a function 
               continuous in $\R_+^2\times[0,1]$, and for an arbitrarily
               fixed $(y_2,y_3)\in\R_+\times[0,1]$ the function $\phi
               (.,y_2,y_3):y_1\to\phi(y_1,y_2,y_3)$ is continuously
               differentiable in $\R_+$, and the following inequalities 
               hold:
\end{description}
\begin{gather}
a_2\ge\phi(y_1,y_2,y_3)\ge a_1   \label{3.4}\\
\phi(y_1,y_2,y_3)+2\,\frac{\di\phi}{\di y_1}\,(y_1,y_2,y_3)y_1\ge a_3
                                 \label{3.5}\\
\Big\vert \frac{\di\phi}{\di y_1}\,(y_1,y_2,y_3)\Big\vert y_1\le a_4,
                                 \label{3.6}
\end{gather}
where $a_i,1\le i\le 4$, are positive numbers.

Inequality \eqref{3.4} 
indicates that the viscosity is bounded from below and from above by positive
constants.
The inequality \eqref{3.5} implies that for fixed values of $\vert E\vert$
and $\mu(u,E)$ the derivative of the function $I(v)\to G(v)$ is positive, where
$G(v)$ is the second invariant of the stress deviator
\begin{equation}
G(v)=4[\phi(I(v),\vert E\vert,\,\,\mu(u,E))]^2I(v).  \notag
\end{equation}
This means that in the case of simple shear flow the shear stress
increases with  increasing shear rate. \eqref{3.6} is a restriction
on $\frac{\di\phi}{\di y_1}$ for large values of $y_1$. 
These inequalities are natural from the physical point of view.

Relative to the function of slip $\chi$, we assume that the following
conditions are satisfied: 

\begin{description} 
\item[(C2)]
           $\chi:(y_1,y_2)\to\chi(y_1,y_2)$ is a function continuous
           in $\R\times\R_+$, and for an arbitrarily fixed $y_1\in\R$,  
           the function $\chi(y_1,.):y_2\to\chi(y_1,y_2)$ is continuously
           differentiable in $\R_+$, and the following inequalities hold:
\end{description}
\begin{gather}
b_2\ge \chi(y_1,y_2)\ge b_1,                                    \label{3.7}\\
\chi(y_1,y_2)+2\frac{\di\chi}{\di y_2}(y_1,y_2)y_2\ge b_3,      \label{3.8}\\
\Big\vert\frac{\di\chi}{\di y_2}(y_1,y_2)\Big\vert y_2\le b_4  \label{3.9},
\end{gather}
where $(y_1,y_2)\in\R\times\R_+$, and $b_i$, $1\le i\le 4$, are positive
numbers. Inequalities \eqref{3.7}--\eqref{3.9} are analogous to the ones
of \eqref{3.4}--\eqref{3.6}. Inequality \eqref{3.7} means that the function
of slip is bounded from below and from above by positive constants.
\eqref{3.8} implies that for fixed value of $y_1$, i.e. the value of
$F_{r\nu}(p,u)(s)$, $s\in S_1$, the derivative of the function 
$\vert u^\tau\vert\to\vert F^\tau\vert=\chi(F_{r\nu}(p,u),\vert u^\tau
\vert^2)\vert u^\tau\vert$ is not less than $b_3$. Indeed, denoting
$z=y_2^{\frac12}$, we obtain
\begin{equation}\notag
\frac{\di(\chi(y_1,z^2)z)}{\di z}=\chi(y_1,z^2)+2\frac{\di\chi}{\di y_2}(y_1,
z^2)z^2\ge b_3,
\end{equation}
that is the frictional force increases as the velocity of slip increases.

The inequality \eqref{3.9} is a restriction on $\frac{\di\chi}{\di y_2}$
for large values of $y_2$. The inequalities \eqref{3.7}--\eqref{3.9} are
natural from the physical viewpoint.

We suppose also that
\begin{equation}\label{3.10}
K=(K_1,\dots,K_n)\in L_2(\Omega)^n, \qquad F=(F_1,\dots,F_n)\in L_2(S_2)^n.
\end{equation}

\section{Boundary value problem.}
We study a problem of searching for a pair of functions $(u,p)$ which 
satisfy the motion equation \eqref{3.1}, the condition of incompressibility
\eqref{3.2} and the boundary conditions \eqref{2.6}, \eqref{2.14} and 
\eqref{3.3}.

Consider the following spaces
\begin{align}
&Z=\{v\,\vert v\in H^1(\Omega)^n,\quad v_\nu\vert_{S_1}=0\},  \label{4.1}\\
&W=\{v\,\vert v\in Z, \quad \diver v=0\}.                    \label{4.2} 
\end{align}

\begin{lem}
Let $\Omega$ be a bounded domain in $\R^n$, $n=2$ or $3$ with a Lipschitz 
continuous boundary $S$, and let $S_1$ be an open nonempty subset of $S$.
Then the expression
\begin{equation}\label{4.3}
\|v\|_Z=\Big(\int_\Omega I(v)\,dx+\int_{S_1}\sum_{k=1}^n v_{\tau k}^2\,ds
\Big)^{\frac12} 
\end{equation}
defines a norm in $Z$ and $W$ being equivalent to the norm of $H^1(\Omega)^n$.
\end{lem}
For a proof see in \cite{7}, Section 1.7. 

Everywhere below we use the following notations: If $Y$ is a normed space,
we denote by $Y^*$ the dual of $Y$, and by $(f,h)$ the duality between $Y^*$
and $Y$, where $f\in Y^*$, $h\in Y$. In particular, if $f\in L_2(\Omega)$ or
$f\in L_2(\Omega)^n$, then $(f,h)$ is the scalar product in $L_2(\Omega)$ or
in $L_2(\Omega)^n$, respectively. The sign $\rightharpoonup$ denotes weak
convergence in a Banach space.

Denote by $B$ the operator of divergence, i.e.
\begin{equation}\label{4.4}
Bu=\diver u.
\end{equation}
It is obvious that $B$ is a linear continuous mapping of $Z$ into $L_2
(\Omega)$, i.e. $B\in \cal L(Z,L_2(\Omega))$. We denote by $B^*$ the adjoint
to $B$ operator.

We introduce operators $M:Z\to Z^*$ and $A:Z\times L_2(\Omega)\to Z^*$ as
follows: 
\begin{align}
&(M(u),h)=2\int_\Omega\phi(I(u),\vert E\vert,\mu(u,E))\eps_{ij}(u)
          \eps_{ij}(h)\,dx, \quad u,h\in Z,        \label{4.5}\\
&(A(u,p),h)=\int_{S_1}\chi\Big(F_{r\nu}(p,u),\sum_{k=1}^n u_{\tau k}^2\Big)
             u_{\tau i}h_{\tau i}\,ds, \quad (u,p)\in Z\times L_2(\Omega),
             \quad h\in Z.                             \label{4.6} 
\end{align}
Consider the problem: find a pair $(u,p)$ such that
\begin{gather}
u\in Z, \qquad p\in L_2(\Omega),                         \label{4.7}\\
(M(u),h)+(A(u,p),h)-(B^*\,p,h)=(K+F,h), \qquad h\in Z,   \label{4.8}\\
(Bu,q)=0, \qquad q\in L_2(\Omega).                       \label{4.9}
\end{gather}
Here we use the notations
\begin{equation}\label{4.10}
(K,h)=\int_\Omega K_ih_i\,dx, \qquad (F,h)=\int_{S_2}F_ih_i\,ds.
\end{equation}
A solution of the problem \eqref{4.7}--\eqref{4.9} will be called a 
generalized solution of the problem  \eqref{3.1}, \eqref{3.2}, \eqref{3.3},
\eqref{2.6} and \eqref{2.14}. Indeed, by use of Green's formula it can be
seen that, if $(u,p)$ is a solution of the problem \eqref{3.1}, \eqref{3.2},
\eqref{3.3}, \eqref{2.6} and \eqref{2.14}, then $(u,p)$ is a solution of
the problem \eqref{4.7}--\eqref{4.9}. On the contrary, if $(u,p)$ is a 
solution of the problem \eqref{4.7}--\eqref{4.9}, then $(u,p)$ is a solution
of the problem \eqref{3.1}, \eqref{3.2}, \eqref{3.3}, \eqref{2.6} and 
\eqref{2.14} in the sense of distributions.
\begin{th}
Let $\Omega$ be a bounded domain in $\R^n$, $n=2$  or $3$, with a Lipschitz 
continuous boundary $S$, and suppose that the conditions $(C1)$, $(C2)$ and
\eqref{3.10} are satisfied. Then there exists a solution of the problem
\eqref{4.7}--\eqref{4.9}.
\end{th}

\section{Auxiliary results.}
We consider  four functions $v_1$, $v_2$, $v_3$, $v_4$ such that
\begin{gather}
v_1\in L_2(\Omega),\quad v_1(x)\ge 0\mbox{ a.e. in }\Omega,                                        \notag\\
v_2\in L_\infty(\Omega),\quad v_2(x)\in[0,1]\mbox{ a.e. in }\Omega, \quad
v_3\in L_2(\Omega), \quad v_4\in H^1(\Omega)^n.          \label{5.1}   
\end{gather}
We set $v=(v_1,v_2,v_3,v_4)$ and define the operator $M_v:Z\to Z^*$ as
follows:
\begin{gather}
(M_v(u),e)=2\int_\Omega\phi(I(u),v_1,v_2)\eps_{ij}(u)\eps_{ij}(e)\,dx
   +\int_{S_1}\chi\Big(F_{r\nu}(v_3,v_4),\sum_{k=1}^n u_{\tau k}^2\Big)
   u_{\tau i}\,e_{\tau i}\,ds,                           \notag\\
   \qquad u,e\in Z.                                      \label{5.2}
\end{gather}
\begin{lem}
Suppose that the conditions $(C1)$, $(C2)$ and \eqref{5.1} are satisfied. Then
\begin{gather}
(M_v(u)-M_v(w),u-w)\ge\mu_1\|u-w\|_Z^2, \qquad u,w\in Z,     \label{5.3}\\
\|M_v(u)-M_v(w)\|_{Z^*}\le\mu_2\|u-w\|_Z, \qquad u,w\in Z,  \label{5.4}
\end{gather}
where
\begin{align}
&\mu_1=\min(2a_1,2a_3,b_1,b_3),                           \notag\\
&\mu_2=2a_2+4a_4+b_2+2b_4.   \label{5.5}
\end{align}
\end{lem}
{\bf Proof.} We present the operator $M_v$ in the form
\begin{gather}
M_v=M_1+M_2,                                                  \label{5.6}\\
(M_1(u),e)=2\int_\Omega\phi(I(u),v_1,v_2)\eps_{ij}(u)\eps_{ij}(e)\,dx, 
                                                              \label{5.7}\\
(M_2(u),e)=\int_{S_1}\chi\Big(F_{r\nu}(v_3,v_4),\sum_{k=1}^n u_{\tau k}^2
\Big)\,u_{\tau i}\,e_{\tau i}\,ds, \quad u,e\in Z.            \label{5.8} 
\end{gather}
Let $u$, $w$ be arbitrary functions in $Z$ and
\begin{equation}\label{5.9}
h=u-w.
\end{equation}
We introduce the function $\gamma$ as follows:
\begin{equation}\label{5.10}
\gamma(t)=\int_\Omega\phi(I(w+th),v_1,v_2)\eps_{ij}(w+th)\eps_{ij}(e)\,dx,
\quad t\in [0,1],\quad e\in Z.
\end{equation}
It is obvious that
\begin{equation}\label{5.11}
\gamma(1)-\gamma(0)=\frac12(M_1(u)-M_1(w),e).
\end{equation}
By using the theorem on the differentiability of a function represented as an integral, we conclude that $\gamma$ is differentiable at any point $t\in(0,1)$. Therefore
\begin{equation}\label{5.12}
\gamma(1)=\gamma(0)+\frac{d\gamma}{dt}(\xi), \quad \xi\in(0,1),
\end{equation}
where
\begin{gather}
\frac{d\gamma}{dt}(\xi)=\int_\Omega[\phi(I(w+\xi h),v_1,v_2)\eps_{ij}
                        (h)\eps_{ij}(e)                   \notag\\
+2\frac{\di\phi}{\di y_1}\,(I(w+\xi h),v_1,v_2)\eps_{km}(w+\xi h)\eps_{km}
          (h)\eps_{ij}(w+\xi h)\eps_{ij}(e)]dx.           \label{5.13}
\end{gather}
Taking note of the inequality 
\begin{equation}\label{5.14a}
\vert\eps_{km}(w+\xi h)\eps_{km}(h)\vert\le I(w+\xi h)^{\frac12}
          I(h)^{\frac12},
\end{equation}
and \eqref{3.4}, \eqref{3.6}, \eqref{5.11}--\eqref{5.13}, we obtain
\begin{gather}\label{5.14}
\|M_1(u)-M_1(w)\|_{Z^*}\le (2a_2+4a_4)\Big(\int_\Omega I(u-w)\,dx\Big)
^{\frac12} \notag\\
\le (2a_2+4a_4)\|u-w\|_Z.
\end{gather}
Define the function $g$ as follows:
\begin{equation}
g(\alpha,x)=\begin{cases}
\frac{\di\phi}{\di y_1}(\alpha,v_1(x),v_2(x)), &\mbox{ if }\frac{\di\phi}
{\di y_1}(\alpha,v_1(x),v_2(x))<0,\\
0, &\mbox{ if }\frac{\di\phi}{\di y_1}(\alpha,v_1(x),v_2(x))\ge 0,
\end{cases}\notag
\end{equation}
where $\alpha\in\R_+$, $x\in\Omega$.

Then, taking $e=h$ in \eqref{5.13} and applying \eqref{3.4}, \eqref{3.5},
\eqref{5.9} and \eqref{5.14a}, we obtain
\begin{gather}
\frac{d\gamma}{dt}(\xi)=\int_\Omega[\phi(I(w+\xi h),v_1,v_2)I(h)
\notag\\
+2\frac{\di\phi}{\di y_1}(I(w+\xi h),v_1,v_2)(\eps_{ij}(w+\xi h)
\eps_{ij}(h))^2]dx\ge \min(a_1,a_3)\int_\Omega I(u-w)\,dx   \label{5.15}
\end{gather}
\eqref{5.11}, \eqref{5.12} and \eqref{5.15} imply
\begin{equation}\label{5.16}
(M_1(u)-M_1(w),u-w)\ge 2\min(a_1,a_3)\int_\Omega I(u-w)\,dx.
\end{equation}
We introduce the function $\gamma_1$ as follows:
\begin{equation}
\gamma_1(t)=\int_{S_1}\chi\Big(F_{r\nu}(v_3,v_4),\sum_{k=1}^n(w_{\tau k}
+th_{\tau k})^2\Big)(w_{\tau i}+th_{\tau i})e_{\tau i}\,ds,     
\quad    t\in[0,1], \quad e\in Z, \notag  
\end{equation}
where $h$ is defined by \eqref{5.9}.

By analogy with the foregoing, we obtain
\begin{gather}
(M_2(u)-M_2(w),e)=\gamma_1(1)-\gamma_1(0)=\frac{d\gamma_1}{dt}(\xi_1)
                                                         \notag\\
   =\int_{S_1}\Big[\chi\Big(F_{r\nu}(v_3,v_4),\sum_{k=1}^n(w_{\tau k}+
\xi_1h_{\tau k})^2\Big) h_{\tau i}\,e_{\tau i}+2\frac{\di\chi}{\di y_2}
\Big(F_{r\nu}(v_3,v_4),\sum_{k=1}^n(w_{\tau k}+\xi_1h_{\tau k})^2\Big)
                                                          \notag\\
    \times(w_{\tau k}+\xi_1h_{\tau k})h_{\tau k}(w_{\tau i}+\xi_1h_
{\tau i})    e_{\tau i}\Big]ds, \qquad \xi_1\in (0,1),    \label{5.18} 
\end{gather}
and \eqref{3.7}--\eqref{3.9} imply
\begin{gather}
\|M_2(u)-M_2(w)\|_{Z^*}\le (b_2+2b_4)\Big(\int_{S_1}\sum_{i=1}^n
(u_{\tau i}-w_{\tau i})^2\,ds\Big)^{\frac12} \notag \\\le(b_2+2b_4)\|u-w\|_Z,  \label{5.19}\\
(M_2(u)-M_2(w),u-w)\ge\min(b_1,b_3)\int_{S_1}\sum_{k=1}^n(u_{\tau k}
-w_{\tau k})^2\,ds.  \label{5.20}
\end{gather}
Taking \eqref{5.14}, \eqref{5.16}, \eqref{5.19} and \eqref{5.20} into
account, we obtain \eqref{5.3}--\eqref{5.5}.
$\blacksquare$

Let $\alpha$ be a positive number. Define the operator $A_\alpha:Z\to
Z^*$ as follows:
\begin{equation}\label{5.21}
(A_\alpha(u),h)=\int_{S_1}\chi\Big(F_{r\nu}(-\frac1{\alpha}\,Bu,u),
\sum_{k=1}^n u_{\tau k}^2\Big)u_{\tau i}h_{\tau i}\,ds \quad u,h\in Z.
\end{equation}
Consider the problem: find a function $u_\alpha$ satisfying
\begin{gather}
u_\alpha\in Z,                                       \label{5.22}\\
(M(u_\alpha),h)+(A_\alpha(u_\alpha),h)+\frac1\alpha(Bu_\alpha,Bh)
=(K+F,h), \qquad h\in Z.                              \label{5.23}
\end{gather}
The problem \eqref{5.22}, \eqref{5.23} is an approximation of the problem
\eqref{4.7}--\eqref{4.9} in which the function of pressure $p$ is replaced by the function
$-\alpha^{-1}\diver u$; in this case we do not  assume that $\diver 
u_\alpha=0$.

\begin{th}

Let $\Omega$ be a bounded in $\R^n$, $n=2$ or $3$, with a Lipschitz 
continuous boundary $S$. Suppose that the conditions $(C1)$, $(C2)$ and
\eqref{3.10} are satisfied. Then for an arbitrary $\alpha>0$, there exists
a solution of the problem \eqref{5.22}, \eqref{5.23}.
\end{th}
{\bf Proof.} Let $\{Z_m\}_{m=1}^\infty$ be a sequence of finite dimensional
subspaces in $Z$ such that
\begin{gather}
\lim_{m\to\infty}\,\inf_{h\in Z_m}\|v-h\|_Z=0, \qquad v\in Z,   \label{5.24}\\
Z_m\subset Z_{m+1}, \qquad m\in\N.                        \label{5.25}
\end{gather}
We seek an approximate solution of the problem \eqref{5.22}, \eqref{5.23}
in the form
\begin{equation}\label{5.26}
u_{\alpha m}\in Z_m, \quad (M(u_{\alpha m}),h)+(A_\alpha(u_{\alpha m}),
h)+\frac1\alpha(Bu_{\alpha m},Bh)=(K+F,h), \qquad h\in Z_m.
\end{equation}
By \eqref{3.4}, \eqref{3.7}, \eqref{3.10} and \eqref{4.3}, we obtain
\begin{gather}
y(h)=(M(h),h)+(A_\alpha(h),h)+\frac1\alpha(Bh,Bh)-(K+F,h)     \notag\\
\ge 2a_1\int_\Omega I(h)\,dx+b_1\int_{S_1}\sum_{i=1}^n h_{\tau i}^2\,ds
-\|K+F\|_{Z^*}\|h\|_Z
\ge\mu_1\|h\|_Z^2-\|K+F\|_{Z^*}\|h\|_Z,                 \label{5.27}
\end{gather}
where $\mu_1=\min(2a_1,b_1)$.

Therefore, $y(h)\ge 0$ for $\|h\|_Z\ge r=\|K+F\|_{Z^*}\,\mu_1^{-1}$. 

From the corollary of Brouwer's fixed point theorem (cf.\cite{2}), it follows
that there exists a solution of \eqref{5.26} with
\begin{equation}\notag
\|u_{\alpha m}\|_Z\le r, \quad \|M(u_{\alpha m})+A_\alpha(u_{\alpha m})
\|_{Z^*}\le c,   \quad m\in\N,
\end{equation}
where the second inequality follows from \eqref{3.4} and \eqref{3.7}.
Therefore, we can extract a subsequence $\{u_{\alpha\eta}\}_{\eta=1}^\infty$ 
such that
\begin{gather}
u_{\alpha\eta}\rightharpoonup u_\alpha \quad \text{in } Z, \label{5.28}\\
u_{\alpha\eta}\to u_\alpha \quad \text{in } L_2(\Omega)^n \text
 { and a.e. in } \Omega,                                      \label{5.29}\\
M(u_{\alpha\eta})+A_\alpha(u_{\alpha\eta})\rightharpoonup \theta \quad
 \text{in } Z^*.                                             \label{5.30}
\end{gather}
Let $\eta_0$ be a fixed positive integer and $h\in Z_{\eta_0}$. Observing
\eqref{5.29}, \eqref{5.30}, we pass to the limit in \eqref{5.26} with $m$
replaced by $\eta$, and obtain
\begin{equation}\label{5.31}
(\theta+\frac1\alpha\,B^*\,Bu_\alpha,h)=(K+F,h), \qquad h\in Z_{\eta_0}.
\end{equation}
Since $\eta_0$ is an arbitrary positive integer, by \eqref{5.24}, we obtain
\begin{equation}\label{5.32}
\theta+\frac1\alpha\,B^*\,Bu_\alpha=K+F  \quad \text{in } Z^*.
\end{equation}
We present the operators $M$ and $A_\alpha$ in the form
\begin{equation}\label{5.33}
M(u)=\tilde M(u,u), \qquad A_\alpha(u)=\tilde A_\alpha(u,u),
\end{equation}
where the operators $(u,v)\to \tilde M(u,v)$ and $(u,v)\to\tilde A_\alpha
(u,v)$ are mappings of $Z\times Z$ into $Z^*$ according to
\begin{align}
&(\tilde M(u,v),h)=2\int_\Omega\phi(I(v),\vert E\vert,\mu(u,E))\eps_{ij}(v)\eps_
   {ij}(h)\,dx,                                        \label{5.34}\\
&(\tilde A_\alpha(u,v),h)=\int_{S_1}\chi\Big(F_{r\nu}(-\frac1\alpha\,
   Bu,u),\sum_{k=1}^n v_{\tau k}^2\Big)v_{\tau i}h_{\tau i}\,ds.
                                                       \label{5.35} 
\end{align}
Denote
\begin{gather}
X_\eta(v)=(\tilde M(u_{\alpha\eta},u_{\alpha\eta})+\tilde A_\alpha
      (u_{\alpha\eta},u_{\alpha\eta})+\frac1\alpha\,B^*\,B(u_
      {\alpha\eta}-v)                                  \notag\\
-\tilde M(u_{\alpha\eta},v)-\tilde A_\alpha(u_{\alpha\eta},v),
u_{\alpha\eta}-v),    \quad v\in Z.                    \label{5.36}   
\end{gather}
By Lemma 5.1, see \eqref{5.3}, we obtain
\begin{equation}\label{5.37}
X_\eta(v)\ge 0, \qquad \eta\in \N, \qquad v\in Z.
\end{equation} 
We have
\begin{gather}
\|\tilde M(u_{\alpha\eta},v)-\tilde M(u_\alpha,v)\|_{Z^*}  \notag\\
 \le 2\Big\{\int_\Omega[\phi(I(v),\vert E\vert,\mu(u_{\alpha\eta},E))
 -\phi(I(v),\vert E\vert,\mu(u_\alpha,E))]^2\,I(v)\,dx\Big\}^{\frac12}.
                                                             \label{5.38}
\end{gather}
(C1), \eqref{5.29}, \eqref{5.38}, and the Lebesgue theorem give
\begin{equation}\label{5.39}
\tilde M(u_{\alpha\eta},v)\to\tilde M(u_\alpha,v) \quad \text{in } Z^*.
\end{equation}
Likewise, we obtain
\begin{equation}\label{5.40}
\tilde A_\alpha(u_{\alpha\eta},v)\to\tilde A_\alpha(u_\alpha,v) \quad
      \text{in } Z^*.
\end{equation}
\eqref{5.26}, \eqref{5.28} and \eqref{5.33} yield
\begin{equation}\label{5.41}
(\tilde M(u_{\alpha\eta},u_{\alpha\eta})+\tilde A_\alpha(u_{\alpha\eta},
u_{\alpha\eta}),u_{\alpha\eta})+\frac1\alpha(Bu_{\alpha\eta},Bu_
{\alpha\eta})                                     
 = (K+F,u_{\alpha\eta})\to (K+F,u_\alpha).   
\end{equation}
By \eqref{5.30} and \eqref{5.32}, we obtain
\begin{equation}\label{5.42}
\lim[(\tilde M(u_{\alpha\eta},u_{\alpha\eta})+\tilde A_\alpha(u_{\alpha
\eta},u_{\alpha\eta}),v)]+\frac1\alpha(Bu_\alpha,Bv)=(K+F,v), \quad v\in Z.
\end{equation}
Observing \eqref{5.39}--\eqref{5.42}, we pass to the limit in \eqref{5.36}.
Then by \eqref{5.37}, we find
\begin{equation}\label{5.43}
(K+F-\tilde M(u_\alpha,v)-\tilde A_\alpha(u_\alpha,v)-\frac1\alpha
B^*\,Bv,u_\alpha-v)\ge 0, \qquad v\in Z.
\end{equation}
We choose $v=u_\alpha-\gamma h$, $\gamma>0$, $h\in Z$, and consider $\gamma
\to 0$. Then, Lemma 5.1, see \eqref{5.2}, \eqref{5.4}, \eqref{5.34} and 
\eqref{5.35}, implies
\begin{equation}\label{5.44}
(K+F-M(u_\alpha)-A_\alpha(u_\alpha)-\frac1\alpha\,B^*\,Bu_\alpha,h)
\ge 0.
\end{equation}
This inequality holds for any $h\in Z$. Replacing $h$ by $-h$ shows that 
equality holds true  in \eqref{5.44}. Therefore, $u_\alpha$ is a solution 
of the problem \eqref{5.22}, \eqref{5.23}. 
$\blacksquare$

We will use also the following  lemma:
   \begin{lem}
    Let $\Omega$ be a bounded domain in $\R^n$, $n=2$ or $3$, with a Lipschitz
    continuous boundary $S$, and let the operator $B\in\cal L(Z,L_2(\Omega))$ 
    be defined by \eqref{4.4}.
    Then, the $\inf$-$\sup$ condition
    \begin{equation}\label{5.45}
    \inf_{\mu\in L_2(\Omega)}\,\,\sup_{v\in Z}\,\,\frac{(Bv,\mu)}{\|v\|_Z\,
    \|\mu\|_{L_2(\Omega)}}\,\,\ge \beta_1>0
    \end{equation}
    holds true. The operator $B$ is an isomorphism from $W^\bot$  onto $L_2
    (\Omega)$, where $W^\bot$ is orthogonal complement of $W$ in $Z$, and the 
    operator $B^*$ that is     adjoint to $B$, is an isomorphism from $L_2
    (\Omega)$ onto the polar  set 
    \begin{equation}\label{5.46}
    W^0=\{f\in Z^*,\,(f,u)=0, \quad u\in W\}.
    \end{equation}
    Moreover,
    \begin{gather}
    \|B^{-1}\|_{\cal L(L_2(\Omega),W^\bot)}\le \frac1{\beta_1}, \label{5.47}\\
    \|(B^*)^{-1}\|_{\cal L(W^0,L_2(\Omega))}\le \frac1{\beta_1}.\label{5.48}
    \end{gather}
    \end{lem}

    For a proof see in \cite{7}, Section 6.1.2.

\section{Proof of Theorem 4.1.}
Let $\{\alpha_i\}$ be a sequence of positive numbers such that $\lim
\alpha_i=0$. Consider the problem: given $\alpha_i$, find 
$u_{\alpha_i}$ satisfying
\begin{align}
&u_{\alpha_i}\in Z,                            \label{6.1}\\
&(M(u_{\alpha_i}),h)+(A_{\alpha_i}(u_{\alpha_i}),h)+\alpha_i^{-1}\,
(B\,u_{\alpha_i},Bh)=(K+F,h), \qquad h\in Z.                     \label{6.2}
\end{align}
The existence of a solution of the problem \eqref{6.1}, \eqref{6.2}
follows from the Theorem 5.1.

Taking $h=u_{\alpha_i}$ in \eqref{6.2}, we obtain
\begin{equation}\label{6.3}
(M(u_{\alpha_i}),u_{\alpha_i})+(A_{\alpha_i}(u_{\alpha_i}),u_{\alpha_i})
+\alpha_i^{-1}\|Bu_{\alpha_i}\|_{L_2(\Omega)}^2        
\le\|K+F\|_{Z^*}\|u_{\alpha_i}\|_Z.                    
\end{equation}
\eqref{3.4}, \eqref{3.7} and \eqref{6.3} imply
\begin{equation}\label{6.4}
2a_1\int_\Omega I(u_{\alpha_i})dx+b_1\int_{S_1}\sum_{k=1}^n u_{\alpha_i 
\tau k}^2\,ds+\alpha_i^{-1}\|Bu_{\alpha_i}\|_{L_2(\Omega)}^2\le\|K+F\|_{Z^*}
\|u_{\alpha_i}\|_Z.
\end{equation}
It follows from here and \eqref{4.3} that
\begin{align}
&\|u_{\alpha_i}\|_Z\le c_1,                                    \label{6.5}\\
&\alpha_i^{-\frac12}\|Bu_{\alpha_i}\|_{L_2(\Omega)}\le c_2.  \label{6.6}
\end{align}
Therefore, a subsequence $\{u_{\alpha_m}\}$ can be extracted from the
sequence $\{u_{\alpha_i}\}$ such that
\begin{align}
&u_{\alpha_m}\rightharpoonup u \quad \text{in } Z,           \label{6.7}\\
&u_{\alpha_m}\to u \quad \text{in } L_2(\Omega)\quad \text{and a.e. in } 
             \Omega,                                          \label{6.8}\\
&u_{\alpha_m}\vert_{S_1}\to u\vert_{S_1} \quad \text{in } L_2(S_1)\quad
             \text{and a.e. in } S_1,                         \label{6.9}\\
&Bu_{\alpha_m}\to 0 \quad \text{in } L_2(\Omega).            \label{6.10}
\end{align}
\eqref{6.2} yields
\begin{equation}\label{6.11}
\alpha_m^{-1}B^*Bu_{\alpha_m}=K+F-M(u_{\alpha_m})-A_{\alpha_m}(u_
{\alpha_m}) \quad \text{in } Z^*.
\end{equation}
By virtue of \eqref{3.4}, \eqref{3.7}, \eqref{6.5} and \eqref{6.9} the 
right-hand side of \eqref{6.11} is bounded in $Z^*$. Therefore, Lemma 5.2,
yields
\begin{equation}\label{6.12}
\alpha_m^{-1}\|Bu_{\alpha_m}\|_{L_2(\Omega)}\le c_3,
\end{equation}
and we can consider that
\begin{equation}\label{6.13}
\alpha_m^{-1}\,Bu_{\alpha_m}\rightharpoonup p \quad \text{in } L_2(\Omega).
\end{equation}
By analogy with the proof of Theorem 5.1, we pass to the limit in \eqref{6.2}
using  \eqref{6.7}--\eqref{6.10} and \eqref{6.13}. As a result, we obtain
that the pair $(u,p)$ is a solution of the problem \eqref{4.7}--\eqref{4.9}.

{\bf Remark.} Suppose that the condition of slip has the following form:
\begin{gather}
2\phi(I(u),\vert E\vert,\mu(u,E))\Big[\eps_{ij}(u)\nu_j-\eps_{kj}(u)\nu_j\nu_k\nu_i\Big]               \notag\\
    =-\chi(F_{r\nu}(p,u),\,\sum_{k=1}^n(Pu_{\tau k})^2)u_{\tau i} \quad
      \text{on } S_1, \quad i=1,\dots,n,                 \label{6.14}
\end{gather}
 compare with \eqref{2.14}. Here $P$ is the operator of regularization defined by \eqref{2.9}, 
\eqref{2.10}.

In this case, the function $\chi$ in the operator $A$ in \eqref{4.6} is defined by just the same expression as in \eqref{6.14}.

The condition of slip \eqref{6.14} is reasonable from the physical point of view and under such condition the Theorem 5.1 remains true without the restrictions \eqref{3.8} and \eqref{3.9}.

\section{Galerkin method for the problem \eqref{4.7}--\eqref{4.9}.}
Let $\{N_m\}$ be a sequence of finite-dimensional subspaces in $L_2(\Omega)$
such that:
\begin{gather}
\lim_{m\to\infty}\,\,\inf_{y\in N_m}\,\|w-y\|_{L_2(\Omega)}=0, \quad 
 w\in L_2(\Omega).                    \label{7.1}\\
N_m\subset N_{m+1}, \qquad m\in\N,    \label{7.2}\\
\inf_{\mu\in N_m}\,\sup_{v\in Z_m}\,\frac{(B_m v,\mu)}{\|v\|_Z\|\mu\|_{L_2
(\Omega)}}\,\ge\beta>0,\quad m\in\N,  \label{7.3}
\end{gather}
where the operators $B_m\in\cal L(Z_m,N_m^*)$ are defined as follows:
\begin{equation}\label{7.4}
(B_m v,\mu)=\int_\Omega \mu\,\diver v\,dx \qquad v\in Z_m, \quad \mu\in N_m.
\end{equation}
   Let $B_m^*\in\cal L(N_m,Z_m^*)$ be the adjoint operator of $B_m$ with
$(B_m v,\mu)=(v,B_m^*\,\mu)$ for all $v\in Z_m$ and all $\mu\in N_m$.

We introduce the spaces $W_m$ and $W_m^0$ by:
\begin{gather}
W_m=\{v\in Z_m,\,\,\,(B_m v,\mu)=0,\quad \mu\in N_m\}, \label{7.5}\\
W_m^0=\{q\in Z_m^*,\,\,\,(q,v)=0, \quad v\in W_m\}.    \label{7.6} 
\end{gather}

\begin{lem}
Let $\{Z_m\}_{m=1}^\infty$, $\{N_m\}_{m=1}^\infty$ be sequences of 
finite-dimensional subspaces in $Z$ and $L_2(\Omega)$ such that \eqref{7.3}
holds true.
Then the operator $B_m^*$ is an isomorphism from $N_m$ onto $W_m^0$, and the
operator $B_m$ is an isomorphism from $W_m^\bot$ onto $N_m^*$, where $W_m^
\bot$ is an orthogonal complement of $W_m$ in $Z_m$. Moreover, 
\begin{equation}\label{7.7}
\|(B_m^*)^{-1}\|_{\cal L(W_m^0,N_m})\le\frac1{\beta}, \quad \|B_m^{-1}\|_
{\cal L(N_m^*,W_m^\bot)}\le \frac1{\beta},\quad  m\in\N.
\end{equation}
\end{lem}
For a proof see in \cite{7}, Section 6.1.2.

We seek an approximate solution of the problem \eqref{4.7}--\eqref{4.9}
of the form:
\begin{gather}
(u_m,p_m)\in Z_m\times N_m,                         \label{7.8}\\
(M(u_m),h)+(A(u_m,p_m),h)-(B_m^*\,p_m,h)=(K+F,h)     \qquad h\in Z_m,
                                                    \label{7.9}\\
(B_m u_m,q)=0, \qquad q\in N_m,                     \label{7.10}
\end{gather}
\begin{th}
Let $\Omega$ be a bounded domain in $\R^n$, $n=2$ or $3$, with a Lipschitz
continuous boundary $S$, and suppose that the conditions $(C1)$, $(C2)$ and
\eqref{3.10} are satisfied. Let also $\{Z_m\}_{m=1}^\infty$ and 
$\{N_m\}_{m=1}^\infty$ be sequences of 
finite-dimensional subspaces in $Z$ and $L_2(\Omega)$ which  satisfy the
conditions \eqref{5.24}, \eqref{5.25}, \eqref{7.1}--\eqref{7.3}. Then for an
arbitrary $m\in\N$, there exists a solution of the problem 
\eqref{7.8}--\eqref{7.10}, and a subsequence $\{u_k,p_k\}$ can be extracted
from the sequence $\{u_m,p_m\}$ such that:
\begin{align}
&u_k\to u \quad \text{in } Z,            \label{7.11}\\
&p_k\to p \quad \text{in } L_2(\Omega),  \label{7.12}
\end{align}
where $(u,p)$ is a solution of the problem \eqref{4.7}--\eqref{4.9}.
\end{th}
{\bf Proof.} 
1) We determine a mapping $A_1:L_2(\Omega)\times Z\times Z \to Z^*$ as
follows:
\begin{gather}
(A_1(\mu,v,w),h)=\int_{S_1}\chi(F_{r\nu}(\mu,v),\,\sum_{k=1}^n\,w_{\tau k}
^2)w_{\tau i}h_{\tau i}\,ds,                         \notag\\
\mu\in L_2(\Omega), \quad  (v,w,h)\in Z^3.           \label{7.13} 
\end{gather}

Consider the following the problem: given a pair $(v_m,\mu_m)\in Z_m\times
N_m$, find $(\hat v_m,\hat\mu_m)$ satisfying
\begin{gather}
(\hat v_m,\hat\mu_m)\in Z_m\times N_m,               \label{7.14}\\
(\tilde M(v_m,\hat v_m),h)+(A_1(\mu_m,v_m,\hat v_m),h)-(B_m^*\,\hat\mu_m,h)
=(K+F,h),  \qquad h\in Z_m,                          \label{7.15}\\
(B_m,\hat v_m,q)=0, \qquad q\in N_m.                 \label{7.16}
\end{gather}
It follows from \eqref{7.5} and \eqref{7.14}--\eqref{7.16} that $\hat v_m$
is a solution of the problem:
\begin{gather}
\hat v_m\in W_m,                      \label{7.17}\\
(\tilde M(v_m,\hat v_m),h)+(A_1(\mu_m,v_m,\hat v_m),h)=(K+F,h) \qquad
h\in W_m.                             \label{7.18}
\end{gather}
By Lemma 5.1, the operator
\begin{equation}\notag
\tilde M(v_m,.)+A_1(\mu_m,v_m,.):v\to\tilde M(v_m,v)+A_1(\mu_m,v_m,v) 
\end{equation}
is a mapping of $Z$ into $Z^*$ that is strictly monotone, coercive and 
continuous. Therefore, there exists a unique solution of the problem
\eqref{7.17}, \eqref{7.18}, and by Lemma 7.1, there exists a unique function
$\hat\mu_m\in N_m$  such that:
\begin{equation}\label{7.19}
B_m^*\hat\mu_m=\tilde M(v_m,\hat v_m)+A_1(\mu_m,v_m,\hat v_m)-K-F \quad
      \text{in } Z_m^*.
\end{equation}
In this case the pair $(\hat v_m,\hat\mu_m)$ is a unique solution of the 
problem \eqref{7.14}--\eqref{7.16}.

We take $h=\hat v_m$ in \eqref{7.18}. Then by \eqref{3.4} and \eqref{3.7},
we obtain
\begin{equation}\label{7.20}
\|\hat v_m\|_Z\le c_1, \quad (v_m,\mu_m)\in Z_m\times N_m, \quad m\in \N.
\end{equation}
Lemma 7.1 and \eqref{7.19}  imply
\begin{equation}\label{7.21}
\|\hat\mu_m\|_{L_2(\Omega)}\le c_2,  \quad (v_m,\mu_m)\in Z_m\times N_m,
\quad m\in\N.
\end{equation}
For $m\in\N$, we introduce a mapping $\cal B_m:Z_m\times N_m\to Z_m\times 
N_m$ as follows: $(v_m,\mu_m)\in Z_m\times N_m$, $\cal B_m(v_m,\mu_m)=
(\hat v_m,\hat\mu_m)$, where $(\hat v_m,\hat\mu_m)$ is the solution of the
problem \eqref{7.14}--\eqref{7.16}.

Let $\{g_k,\alpha_k\}\in Z_m\times N_m$ and $g_k\to g$, $\alpha_k\to\alpha$.
By using (C1), (C2), \eqref{7.20}, and \eqref{7.21}, one can verify that
$\cal B_m(g_k,\alpha_k)\to \cal B_m(g,\alpha)$. Hence, $\cal B_m$ is a continuous mapping
of $Z_m\times N_m$ into itself.

Moreover, \eqref{7.20} and \eqref{7.21} yield that the mapping $\cal B_m$
maps a compact convex set
\begin{equation}\notag
d_m=\{(v,\mu)\in Z_m\times N_m, \quad \|v\|_Z\le c_1, \quad \|\mu\|_
{L_2(\Omega)}\le c_2\}
\end{equation}
into itself. Therefore, the Schauder principle implies that there exists a
pair $(u_m,p_m)\in Z_m\times N_m$ such that:
\begin{equation}\label{7.22}
\cal B_m(u_m,p_m)=(u_m,p_m).
\end{equation}
In addition, the pair $(u_m,p_m)$ is a solution of the problem 
\eqref{7.8}--\eqref{7.10} for any $m$, and we have
\begin{equation}\label{7.23}
\|u_m\|_Z\le c_1, \qquad \|p_m\|_{L_2(\Omega)}\le c_2, \qquad m\in\N.
\end{equation}
Hence, a subsequence $\{u_k,p_k\}$ can be extracted from the sequence $\{
u_m,p_m\}$ such that:
\begin{align}
&u_k\rightharpoonup u_0 \quad \text{in } Z,                    \label{7.24} \\
&u_k\to u_0 \quad \text{in } L_2(\Omega) \quad \text{and a.e. in } \Omega, 
                                                               \label{7.25} \\
&p_k\rightharpoonup p_0 \quad \text{in } L_2(\Omega),          \label{7.26} \\
&F_{r\nu}(p_k,u_k)\to F_{r\nu}(p_0,u_0) \quad \text{in } L_\infty(S_1),
                                                               \label{7.27} \\
&M(u_k)+A(u_k,p_k)\rightharpoonup \Theta \quad \text{in } Z^*. \label{7.28} 
\end{align}
Let $k_0$ be a fixed positive integer and let $h\in Z_{k_0}$, $q\in N_{k_0}$.
By \eqref{7.24}, \eqref{7.26}, \eqref{7.28}, we pass to the limit in 
\eqref{7.9}, \eqref{7.10} with $m$ changed by $k$, which gives
\begin{gather}
(\Theta-B^*\,p_0, h)=(K+F,h),  \qquad   h\in Z_{k_0},    \label{7.29}\\
\int_\Omega q\diver u_0\,dx=0, \qquad   q\in N_{k_0}.    \label{7.30}
\end{gather}
Since $k_0$ is an arbitrary positive integer, we obtain by \eqref{5.24}, 
\eqref{7.1},  \eqref{7.29} and \eqref{7.30} that
\begin{gather}
\Theta-B^*\,p_0=K+F,   \quad \text{in } Z^*             \label{7.31}\\
\diver u_0=0.                       \label{7.32}
\end{gather}
We determine a mapping $J_k:Z\to Z^*$ as follows:
\begin{equation}\label{7.33}
(J_k(v),h)=(\tilde M(u_k,v)+A_1(p_k,u_k,v),h), \qquad k=0,1,2,\dots .
\end{equation}
It follows from \eqref{5.34} and \eqref{7.13} that
\begin{equation}\label{7.34}
J_k(u_k)=M(u_k)+A(u_k,p_k),
\end{equation}
and Lemma 5.1 gives
\begin{equation}\label{7.35}
(J_k(u_k)-J_k(v),u_k-v)\ge 0, \qquad v\in Z, \quad k=0,1,2,\dots .
\end{equation}
\eqref{7.25}, \eqref{7.27} and the Lebesgue theorem imply
\begin{align}
&\lim(J_k(v),u_k)=(J_0(v),u_0),    \notag\\
&\lim(J_k(v),v)=J_0(v),v),       \label{7.36}
\end{align}
By \eqref{7.28}, \eqref{7.31} and \eqref{7.34}, we have
\begin{equation}\label{7.37}
\lim(J_k(u_k),v)-(B^*\,p_0,v)=(K+F,v).
\end{equation}
Taking into account that
\begin{equation}\notag
(B_k^*\,p_k,u_k)=(p_k,Bu_k)=0,
\end{equation}
we get from \eqref{7.9}, \eqref{7.24} and \eqref{7.34} that
\begin{equation}\label{7.38}
(J_k(u_k),u_k)=(K+F,u_k)\to (K+F,u_0).
\end{equation}
Upon \eqref{7.36}--\eqref{7.38}, we pass to the limit in \eqref{7.35},
which gives
\begin{equation}\label{7.39}
(K+F-J_0(v)+B^*\,p_0,u_0-v)\ge 0, \qquad v\in Z.
\end{equation}
Take here $v=u_0-\xi h$, $\xi>0$, $h\in Z$, and let $\xi$ tend to zero.
By Lemma 5.1, see \eqref{5.4}, we get
\begin{equation}\label{7.40}
(K+F-J_0(u_0)+B^*\,p_0,h)\ge 0, \qquad h\in Z.
\end{equation}
Therefore, the pair $u=u_0$, $p=p_0$ is a solution of the problem 
\eqref{4.7}--\eqref{4.9}.

2) We will show that the solution of the problem \eqref{7.8}--\eqref{7.10}
converge to the solution of \eqref{4.7}--\eqref{4.9} strongly.

Let
\begin{equation}\label{7.41}
Y_k=(J_k(u_k)-J_0(u_0),u_k-u_0).
\end{equation}
Obviously
\begin{equation}\label{7.42}
Y_k=(J_k(u_k)-J_k(u_0),u_k-u_0)+(J_k(u_0)-J_0(u_0),u_k-u_0).
\end{equation}
Upon \eqref{7.24}, \eqref{7.36}--\eqref{7.38} and \eqref{7.41} $\lim Y_k=0$.
By \eqref{7.24}, \eqref{7.36} the second addend in \eqref{7.42} tends also
to zero, and so
\begin{equation}\label{7.43}
\lim (J_k(u_k)-J_k(u_0),u_k-u_0)=0.
\end{equation}
Observing \eqref{7.33}, \eqref{7.43} and Lemma 5.1 see 
\eqref{5.3}, we obtain \eqref{7.11}.

We take $h\in Z_k$ in \eqref{4.8} and \eqref{7.9} and subtract \eqref{4.8}
from \eqref{7.9}. Then, we obtain
\begin{gather}
(B^*(p_k-\mu),h)=(M(u_k)+A(u_k,p_k)-M(u_0)-A(u_0,p_0),h)+(B^*(p_0-\mu),h),
\notag\\
     h\in Z_k, \quad \mu\in N_k.  \label{7.44}
\end{gather}
This equality together with \eqref{7.3} yields
\begin{equation}\label{7.45}
\|p_k-\mu\|_{L_2(\Omega)}\le \sup_{h\in Z_k}\frac{(B^*(p_k-\mu),h)} 
{\beta\|h\|_Z}\le\beta^{-1}\,\cal A_k+c\|p_0-\mu\|_{L_2(\Omega)}, \qquad
\mu\in N_k,  
\end{equation}
where
\begin{equation}\label{7.46}
\cal A_k=\|M(u_k)+A(u_k,p_k)-M(u_0)-A(u_0,p_0)\|_{Z^*}.
\end{equation}
Hence
\begin{gather}
\|p_0-p_k\|_{L_2{\Omega}}\le\inf_{\mu\in N_k}(\|p_0-\mu\|_{L_2(\Omega)})
+\|p_k-\mu\|_{L_2(\Omega)})
\notag\\
\le\beta^{-1}\,\cal A_k+(c+1)\inf_{\mu\in N_k}\|p_0-\mu\|_{L_2(\Omega)}.
\label{7.47}
\end{gather}
By \eqref{7.11} and \eqref{7.26}, we obtain $\lim \cal A_k=0$, and  
\eqref{7.1} implies
\begin{equation}\notag
\lim_{k\to\,\infty}\,\inf_{\mu\in N_k}\|p_0-\mu\|_{L_2(\Omega)}=0,
\end{equation}
so that \eqref{7.12} follows from \eqref{7.47}.
$\blacksquare$


\begin{thebibliography}{99}

\bibitem{1} Belonosov M.S., Litvinov W.G., Finite element method for 
nonlinearly viscous fluids, Z.angew. Math. Mech. {\bf 76}, 307--320, 1996.

\bibitem{2} Gajewski H., Gr\"oger K., Zacharias, Nichtlineare 
Operatorgleichungen und Operatordifferentialgleichungen, Akademie-Verlag,
Berlin, 1974.

\bibitem{3} Hoppe R.H.W., Litvinov W.G., Problems on electrorheological
fluid flows, Preprint, Institute of Mathematics, University of Augsburg,
2001.

\bibitem{4} Ladyzhenskaya O., Solonnikov V., Some problems of vector 
analysis and generalized formulation of boundary value problems for the
Navier-Stokes equations, Zap. Nauchn. Sem. Leningrad. Otdel Math. Inst.
Steklov (LOMI), {\bf 59}, 81--116, 1976 (in Russian).

\bibitem{5} Lemaire E., Bossis G., Yield stress and wall effects in magnetic
colloidal suspensions, J. Phys. D: Appl. Phys., {\bf 24}, 1473--1477, (1991).

\bibitem{6} Lions J.-L., Quelques M\'ethodes de R\'esolution des Probl\`emes
aux Limites non Lin\'eaires, Dunod Gauthier-Villars, Paris, 1969.

\bibitem{7} Litvinov W.G., Optimization in Elliptic Problems with Applications
to Mechanics of Deformable Bodies and Fluid Mechanics, Birkh\"auser, 2000.

\bibitem{8} Litvinov W.G., Motion of Nonlinearly Viscous Fluid, Moscow,
Nauka,1982 (in Russian).

\bibitem{9} Parthasarathy M., Kleingenberg D.J., Electrorheology: mechanisms,
and models, Material Science and Engineering, R17, 57--103, 1996.



\end{thebibliography}
\end{document}